\documentclass[a4paper]{panl}
\usepackage{cite}
\usepackage{wrapfig}
\usepackage{graphicx}
\usepackage{amssymb}
\usepackage{amsfonts}
\usepackage{amsmath}
\usepackage{longtable}
\usepackage{rotating}
\usepackage{lscape}
\usepackage{epsfig}
\usepackage{multirow}
\usepackage{caption}

\def\Mhexp{\ensuremath{M_h^{\mathrm{exp}}}}
\def\GeV{\ensuremath{\mathrm{GeV}}}
\def\Veff{\ensuremath{V_{\mathrm{eff}}}}
\def\MSbar{\ensuremath{\overline{\mathrm{MS}}}}
\def\mucrit{\ensuremath{\mu^{\mathrm{crit}}}}
\def\Mtcrit{\ensuremath{M_t^{\mathrm{crit}}}}
\def\Mhcrit{\ensuremath{M_h^{\mathrm{crit}}}}
\originalTeX
\begin{document}

\title{An advanced precision analysis of the SM vacuum stability}
\maketitle
\authors{A.V.\,Bednyakov$^{a,b,}$\footnote{E-mail: bednya@theor.jinr.ru \\ Talk given at 
the International Session-Conference of SNP PSD RAS ``Physics of Fundamental Interactions'', JINR, Dubna, 2016}} 
\from{$^{a}$\,Joint Institute for Nuclear Research, 141980, Dubna, Russia}
\vspace{-3mm}
\from{$^{b}$\,Dubna State University, 141982, Dubna, Russia}

\begin{abstract}

	Доклад посвящен проблеме стабильности электрослабого вакуума в Стандартной модели фундаментальных взаимодействий.
	В качестве инструмента исследования рассматривается эффективный потенциал поля Хиггса, который при учете радиационных поправок
	может обладать дополнительным, более глубоким минимумому. 
	Обсуждаются различные методы и приближения, используемые для вычисления эффективного потенциала. 
	Особое внимание уделяется ренормгрупповому подходу, позволившему провести анализ стабильности на трехпетлевом уровне точностью.	
	С помощь явно калибровочно-инвариантной процедуры находятся ограничения на наблюдаемые массы бозона Хиггса и топ-кварка, совместные с требованием абсолютной стабильности
	CМ. Демонстрируется важность учета высших поправок теории возмущений. 
	Также обсуждается потенциальная метастабильность СМ и модификации анализа при учете Новой физики.
\vspace{0.2cm}

	The talk is devoted to the problem of stability of the Standard Model vacuum. 
	The effective potential for the Higgs field, which can potentialy exhibit additional, 
	deeper minimum, is considered as a convenient tool for addressing the problem.
	Different methods and approximations used to calculate the potential are considered.
	Special attention is paid to the renomalization-group approach that allows one to carry out  
	three-loop analysis of the problem. 
	By means of an explicit gauge-independent procedure the absolute stability bounds on the observed Higgs and top-quark masses are derived.
	The importance of high-order corrections is demonstrated.
	In addition, potential metastablity of the SM is discussed together with modifications of 
	the analysis due to some New Physics. 
	 
\end{abstract}
\vspace*{6pt}

\noindent


The Standard Model, although being established as a quantum-field theory in mid 70s of the last century, turns out to be a perfect description 
of many phenomena at scales accessible to current accelerator experiments. The Lagrangian of the model 
\begin{eqnarray}
	\mathcal{L}_{\mathrm{SM}} & = & \mathcal{L}_{\mathrm{Gauge}}(g_1,g_2,g_s) + 
				        \mathcal{L}_{\mathrm{Yukawa}}(Y_u,Y_d,Y_l) + 
				        \mathcal{L}_{\mathrm{Higgs}}(\lambda,m^2)  
	\label{eq:LSM}
\end{eqnarray}
depends on the gauge couplings $g_s,g_2,g_1$, which sets the strength of $SU(3)\times SU(2)\times U(1)$ gauge interactions,
(matrix) Yukawa couplings $Y_{u(d)}$ and $Y_l$ for the up(down)-type quarks and leptons coupled to the Higgs doublet $\Phi$,
and the parameters of the tree-level Higgs potential 
\begin{eqnarray}
	V(\Phi) = m^2 |\Phi|^2 + \lambda |\Phi|^4, \qquad \Phi = \begin{pmatrix} \phi^+ \\ \frac{1}{\sqrt 2}(\phi + i \chi) \end{pmatrix}.
	\label{eq:Vtree}
\end{eqnarray}
	For $m^2<0$ the potential has degenerate minima, which are characterized by non-zero Higgs field vacuum expectation value (vev) $|\langle \Phi \rangle|^2 \equiv v^2/2 \neq 0$. As usual it is a signal of the spontaneous symmetry breaking. 
	Due to the degeneracy one can choose the vacuum, for which only $\phi$ component has non-zero vev $\langle \phi \rangle = v$.
	The latter can be expressed in terms of the Lagrangian parameters (in the leading order) as a space-time independent solution of the classical equation of motion
\begin{eqnarray}
	\left.\frac{\partial V(\phi)}{\partial \phi}\right|_{\phi = v} = 0 \Rightarrow v^2 = -\frac{m^2}{\lambda} 
	\label{eq:vev_tree}
\end{eqnarray}
	Due to interactions with the Higgs-field condensate $\langle \phi \rangle$, elementary particles acquire masses proportional to the corresponding coupling. 
	It is this fact that allows one, given the SM relation $M_h^2 = 2 \lambda v^2$, to deduce the value of the self-coupling $\lambda$ 
	from the Higgs mass $\Mhexp = 125.09(24)$ GeV, measured at the LHC. 

	The obtained value turns out to be rather special. 
	Before the discovery of the Higgs boson both theorist and experimentalist were eagerly searching for hints of New Physics (NP). One way to address this issue
	is to analyze self-consistency of the SM by studying high-energy behavior of the relevant effective couplings. 

\begin{wrapfigure}[19]{r}{0.5\textwidth}
	\centering
    \includegraphics[width=0.48\textwidth]{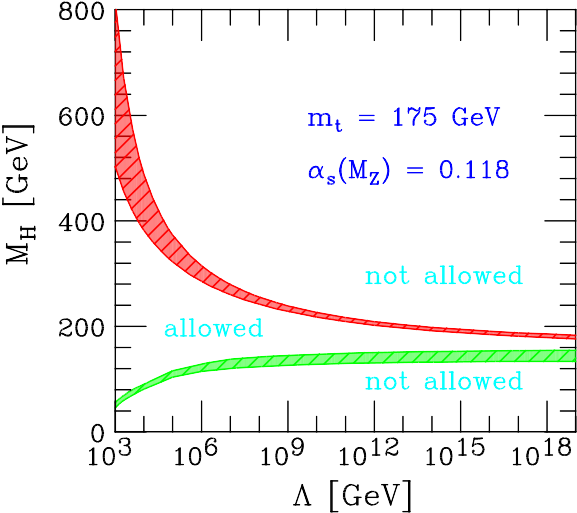}
  \caption{Triviality (upper) and stability (lower) bounds on the Higgs mass $M_h$ depending on the New Physics scale $\Lambda$~\cite{Hambye:1996wb}}
  \label{fig:higgs_mass_constraints}
\end{wrapfigure}

	In Fig.~\ref{fig:higgs_mass_constraints} quite an old plot \cite{Hambye:1996wb} with two curves is presented in the plane $(M_h, \Lambda)$,
	The upper (red) line corresponds to the ``triviality'' constraint - at the corresponding scale $\Lambda$ the self-coupling hits a Landau pole/becomes non-perturbative.
	It is easy to see that the observed value of the Higgs mass lies significantly below this line in the whole range of $\Lambda$. 

	The second (green) curve represents a lower bound on the mass originating from the fact that the Higgs potential \eqref{eq:Vtree} with scale-dependent $\lambda$ 
	becomes unstable for a given scale $\Lambda$, leading to possible decay of the SM vacuum. 
	The curve looks more interesting since $\Mhexp$ lies in a dangerous vicinity of the line. 
	To decide whether the SM can be extrapolated self-consistently up to the Planck scale or New Physics should be introduced to cope with 
	the above-mentioned instability one has to perform an elaborated analysis taking into account various radiative corrections.

	A proper way to study symmetry breaking in the SM is to consider effective potential $\Veff(\phi)$, which also takes into account 
	contribution $\Delta V$ from vacuum fluctuations of all fields of the model
\begin{equation}
		\Veff(\phi)=V(\phi) + \Delta V(\phi) \equiv  
		m^2_{\mathrm{eff}}(\phi) \frac{\phi^2}{2} + \lambda_{\mathrm{eff}}(\phi) \frac{\phi^4}{4},\quad 
		\Delta V(\phi) = \sum\limits_n \hbar^n \Delta^{(n)} V(\phi),
	\label{eq:eff_pot_gen}
\end{equation}
	where $m_\mathrm{eff}$ and $\lambda_{\mathrm{eff}}$ are introduced. 
	For $\Delta V$ loop expansion is implied \cite{Coleman:1973jx,Jackiw:1974cv}. 
	In the Landau gauge we have\footnote{The integral is obviously divergent and can be defined within some renormalization scheme.}    
\begin{eqnarray}
\Delta^{(1)} V (\phi) &=& 
\int \frac{d^4\,k}{2\, (2\pi)^4} \mathrm{STr}\ln \left(k^2 + M^2(\phi)\right),		
		\label{eq:Veff1}
\end{eqnarray}
where field-dependent particle masses $M^2(\phi) = \kappa \phi^2 + \kappa'$ are introduced (see Table~\ref{tab:SM-FD-Mass}).
\begin{table}[t!]
\centering
\begin{tabular}{|c|c|c|c|}
\hline
~~Particle~~& $\kappa$          &~~$\kappa'$~~&~~~n~~~~\\ \hline
$W^\pm$ & $g_2^2/4$         & 0 & $2 \times 3$  \\
$Z$     &~~~$(g_2^2+g_1^2)/4$~~~& 0 & 3 \\
$t$     & $y_t^2/2$   & 0 & $4 \times 3$ \\
$h$     & $3 \lambda$       & $m^2$ & 1 \\
$G$     & $ \lambda $       & $m^2$ & $3 \times 1$ \\
\hline
\end{tabular}
\caption{Field-dependent masses of the SM particles 
$M^2(\phi) = \kappa \phi^2 + \kappa'$  together with the corresponding
numbers of degrees of freedom $n$.
Massive vector bosons $W^+$, $W^-$, and $Z$, top-quark $t$, Higgs boson $h$ and Goldstone 
bosons $G=G^{\pm,0}$ are considered. 
}
\label{tab:SM-FD-Mass}
\end{table}
	The supertrace $\mathrm{STr}$ counts positively (negatively) bosonic (fermionic) degrees of freedom $n$.
	From \eqref{eq:Veff1} one can clearly see the origin of instability \cite{Krasnikov:1978pu} --- the fermionic contribution drives the potential to negative values. 
	High-order terms in the expansion \eqref{eq:eff_pot_gen} can be found from vacuum graphs involving $M^2(\phi)$ in propagators.

	Before going further, let us mention an annoying subtlety of 
	gauge-dependence of the effective potential \cite{Jackiw:1974cv} at general values of $\phi$. 
	It manifests itself as
	a dependence on auxiliary gauge-fixing parameters $\xi_i$.
	The latter is controlled by the following Nielsen identity \cite{Nielsen:1975fs}: 
	\begin{equation}
	\xi \frac{\partial \Veff}{\partial \xi} = - C(\phi,\xi) \frac{\partial \Veff}{\partial \phi},	
		\label{eq:nielsen_Veff}
	\end{equation}
	which tells us that the change due to $\xi$ variation in $\Veff$ can be compensated by appropriate field rescaling.
	It turns out that only at extrema\footnote{Having in mind that $\Veff'(\phi) \equiv J_\phi$, the external source $J_\phi$ vanishes at extrema.} the corresponding energy-density is gauge-invariant
	. 
	We should keep this in mind, when discussing the stability issue of the SM.

	In order to address  the problem of the SM stability, we consider how the values of $\Veff$ at \emph{extrema}
are related to each other. Denoting higgs field value at possible additional minimum as $v'$, we distinguish the case 
of absolute stability of the EW vacuum, $\Veff(v)< \Veff(v')$. 
When $\Veff(v) = \Veff(v')$ we have \emph{critical} situation with degenerate minima
\begin{equation}
	\left.\frac{\partial \Veff}{\partial \phi} \right|_{\phi=v} 
	= \left.\frac{\partial \Veff}{\partial \phi} \right|_{\phi=v'} = 0, \qquad \Veff(v) = \Veff(v'). 	
	\label{eq:critical_conditions}
\end{equation}
Finally, for $\Veff(v)>\Veff(v')$ our ground state turns out to be unstable.
The latter case deserves further study, since there exist a possibility for our Universe to have negligible 
probability to decay into the true vacuum (``metastability''). In what follows we mostly discuss absolute stability and just give a brief comment on potential metastablity.    

\begin{wrapfigure}[19]{r}{0.5\textwidth}
\centering
\vspace{-10pt}
    \includegraphics[width=0.48\textwidth]{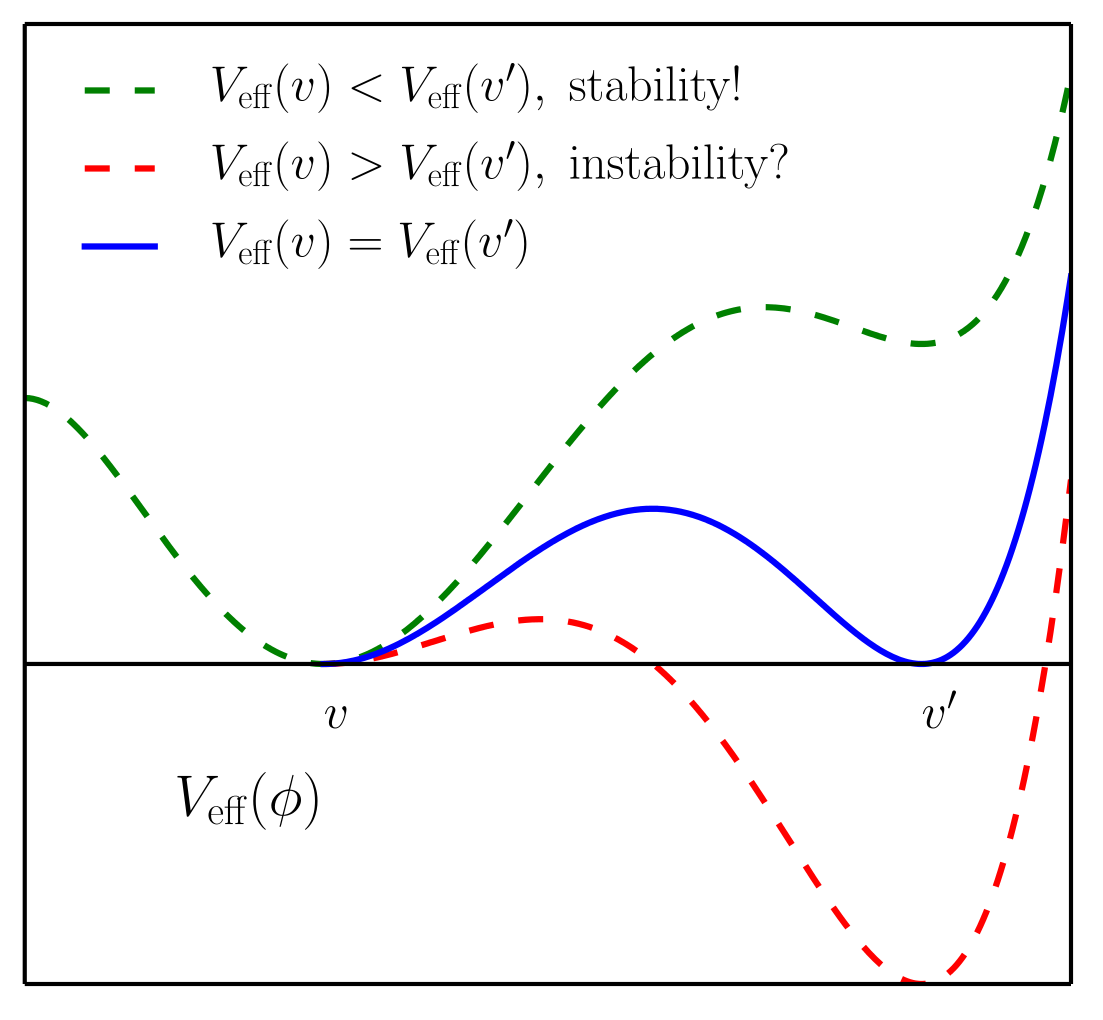}
\vspace{-10pt}
  \captionof{figure}{Different behavior of the effective potential. To decide, whether the EW vacuum with $\langle\phi\rangle=v$ is stable or not,
  one compares $\Veff(v)$ with $\Veff(v')$.}
\label{fig:veff_phases}
\end{wrapfigure}

	It turns out that for large field values $\phi \gg v$ 
	one can not use truncated perturbation-theory (PT) series and has to 
	reorganize (improve) the expansion. 
	In a nutshell, the problem stems from the fact that beyond the tree-level approximation terms of the following form 
\begin{eqnarray}
	a^{L+1}(\mu) \left[ \ln \frac{\phi^2}{\mu^2} \right]^\kappa, \qquad 
	\kappa = 1,\dots,L  
	\label{eq:Logs}
\end{eqnarray}
will arise in $L$-loop contribution to $\Veff$. In \eqref{eq:Logs} $\mu$ corresponds 
to some normalization scale\footnote{We use $\MSbar$ to define the SM running parameters.}, at which a  set of scale-dependent (running) couplings $a(\mu) = \{a_i(\mu)\}$, 
\begin{equation}
	(4\pi)^2 a_i =  \left\{ g_1^2, g_2^2, g_s^2, y_b^2,y_t^2,\lambda\right\},
	\label{eq:couplings}
\end{equation}
is defined.  
The truncated finite-order result \eqref{eq:eff_pot_gen} has residual dependence on $\mu$ and, obviously, has limited precision for $\phi\gg \mu$. 
A well-known solution to this kind of problems is re-summation by means of renormalization group (RG), which, roughly speaking, 
corresponds to the choice $\mu  = \phi$. 

Further simplification can be achieved, when for large field values the full effective potential 
is approximated by the tree-level $\lambda \phi^4$ term 
\begin{eqnarray}
	\Veff(\phi) \stackrel{\phi\gg v}{\Rightarrow} \lambda_{\mathrm{eff}}(\phi) \frac{\phi^4}{4} 
	\simeq \lambda(\mu=\phi)\frac{\phi^4}{4},
	\label{eq:veff_approx}
\end{eqnarray}
with running $\lambda(\mu=\phi)$ in $\MSbar$ scheme \cite{Ford:1992pn}. 
In the latter case \emph{only} 
leading logarithmic (LL) contribution ($\kappa=L$) from \eqref{eq:Logs} can be re-summed consistently. 
Applying the critical conditions \eqref{eq:critical_conditions} to this simple case one obtains
 \cite{Froggatt:1995rt}
\begin{equation}
	\lambda(\mu) = 0, \qquad  \beta_\lambda(\mu) = 0	
	\label{eq:crit_cond_simple}
\end{equation}
	with $(4\pi)^2 \beta_\lambda \equiv d \lambda/d \ln \mu^2$ corresponding to 
	the beta-function of $\lambda$ (see below). 
As independent variables in \eqref{eq:crit_cond_simple} one usually chooses
$\mu$ and a physical mass of interest, either $M_h$ or $M_t$.
Keeping all other parameters fixed, critical scale $\mucrit$ and 
critical mass $\Mhcrit$ ($\Mtcrit$) are deduced.
From \eqref{eq:Veff1} it is easy to convince oneself that the obtained $\Mtcrit$ corresponds to the upper bound 
on the physical mass, while $\Mhcrit$  --- to the lower bound. 

Strictly speaking, if one wants to go beyond the LL approximation, Eq.~\eqref{eq:critical_conditions} should be used in place of \eqref{eq:crit_cond_simple}
and non-logarithmic (``finite'') terms in $\lambda_{\mathrm{eff}}$ should be considered.
The latter are known at two loops \cite{Ford:1992pn,Martin:2001vx} 
within the full SM\footnote{Partial three- and even four-loop results are also known in Landau gauge \cite{potHO} 
.}. 
This allows one to consistently re-summ 
next-to-next-to-leading (NNLL) logarithmic terms ($\kappa = L-2$) 
by utilizing the system of coupled RG equations at the three-loop order

\begin{equation}
	\mu^2 \frac{d a_i}{ d \mu^2} = \beta_{a_i}(a_j),\qquad \beta_{a_i} = \beta^{(1)}_{a_i} + \beta^{(2)}_{a_i} + \beta^{(3)}_{a_i} + ....
	\label{eq:rge_couplings}
\end{equation}
	The appropriate boundary conditions are usually supplied at the EW scale $M_Z \simeq 100$ GeV.
	Various $\beta_a^{(i)}$ represent $i$-loop contribution to the beta-function of $a$ and are known for 
	the full SM up to the third order (see \cite{betas} 
	and refs. therein).
	Some leading four-loop corrections to the beta-functions 
	were recently computed in the literature~\cite{new_betas}. 

	The boundary values are obtained by the so-called matching procedure, which relate the Lagrangian parameters (in the \MSbar~scheme) to a set of (pseudo)observables. 
It is customary to use the pole masses of the SM bosons,  $M_W$, $M_Z$, $M_h$,  
together with that of fermions $M_f$ to extract (or match) the relevant running parameters 
at the EW scale. In principle, matching can be done at any scale. 
	 However, again, 
	 truncated series exhibits bad behavior if $\mu$ is chosen far away from a typical scale 
	 involved in the definition of the observable,  
	 so matching is usually carried out at $\mu\simeq M_Z$ and RG is used for $\mu\gg M_Z$.

	Two more constraints are required in order to completely determine the values of the SM parameters in the \MSbar~scheme. They usually come from the requirement that the SM should reproduce Fermi theory and QCD with five active flavours, when  considered at scales far below the corresponding thresholds 
	(e.g., $M_W$ or $M_t$). In other words, given the SM one can predict the couplings $G_F$ and $\alpha_s^{(5)}(\mu)$ of the above-mentioned effective theories. 
	At the end of the day, we have the relations of the form (full set can be found, e.g., in Ref.\cite{Bednyakov:2015sca})
\begin{eqnarray}
 M^2_h   =  2\lambda v^2 (1 + \bar \delta_h),\quad 
  (4 \pi)^2 \alpha^{(5)}_s(\mu)  =  g_s^2 (1 + \bar \delta \alpha_s), \quad
2^{1/2} G_F  =  v^{-2} (1 + \bar \delta r), 
  \label{eq:matching}
\end{eqnarray}
in which RHS depend only on \MSbar~parameters 
and the renormalization scale $\mu$, while LHS, with the only exception of $\alpha_s^{(5)}(\mu)$, are
$\mu$-independent. Various $\bar \delta$'s represent radiative corrections and are calculated in PT. 
As in the case of effective potential, for consistent NNLO re-summation 
at least two-loop corrections 
in $\bar\delta$ have to be included. 

An important question should be raised when one deals with Eqs.~\eqref{eq:matching}: 
How do we define running vev $v$ entering these equations? 
One option is to assume that $v(\mu)$ is nothing else but the tree-level 
vev \eqref{eq:vev_tree}  and, thus, gauge-independent.
Another common approach utilizes gauge-dependent solution $\phi=\tilde v$ of the equation
\begin{equation}
	\left.\frac{\partial \Veff(\phi,a(\mu),\mu,\xi)}{\partial \phi}\right|_{\phi=\tilde{v}} = 0.
		\label{eq:veff_extrema}
\end{equation}

Both approaches have advantages and disadvantages. Let us mention some of them. 
First of all, particle pole masses are usually identified with the physical ones. 
Due to this, LHS of \eqref{eq:matching} are independent of gauge-fixing parameters $\xi$. 
If one treats $v$ as the tree-level vev, all $\xi$-dependence in RHS is explicit. 
The cancellation is maintained \emph{at the bare level} order-by-order by the inclusion of diagrams with tadpoles
attached to \emph{every} particle coupled to the Higgs boson. 
The drawback of the approach is that the tadpole contributions, 
which typically scale as powers of $M_t^4/(M_W^2 M_h^2)$, may spoil numerical ``convergence'' of PT series.

In the ``tadpole-free'' scheme based on $\tilde v$ \eqref{eq:veff_extrema} the running \MSbar~masses become implicitly 
gauge-dependent and it is very hard to check the necessary cancellation in \eqref{eq:matching}. 
Computations in this case are usually carried out in the Landau gauge $\xi=0$. 
The advantage of this scheme lies in the fact the numerically dangerous tadpole terms are effectively 
re-summed in $\tilde v$ and explicitly contribute only in the Higgs sector 
(see, e.g., Ref.\cite{Actis:2006ra}).

In our study we made use of the first approach. Let us mention, however, that both options should converge to the same result when vev is consistently traded for the Fermi constant $G_F$ by inverting
the last relation of \eqref{eq:matching}. 
This option was advocated, e.g., in Ref.~\cite{mY} 
and renders explicitly 
gauge-independent relations \eqref{eq:matching} for particle masses. 

The set of non-linear equations \eqref{eq:matching} on the Lagrangian parameters can be solved analytically 
in PT: 
\begin{eqnarray}
	\lambda(\mu) = 2^{-1/2} G_F M_H^2 [ 1 + \delta_H(\mu) ], \quad
  y_f(\mu) = 2^{3/4} G_F^{1/2} M_f [ 1 + \delta_f(\mu) ], 
\label{eq:inv_matching}
\end{eqnarray}
	where $\delta_i$ differ from $\bar \delta_i$ and depend on physical masses instead of the running ones.

Due to lack of space, we are not going to present the results for 
the SM couplings at the EW scale 
but just refer to 
our paper \cite{Bednyakov:2015sca} together with the C++ code~\cite{Kniehl:2016enc}.
In spite of the fact that different approaches to tadpoles were utilized in Ref.~\cite{degrassi} 
and our work, 
numerical results turns out to be very close
to each other for the same input taken from PDG2014~\cite{Agashe:2014kda}. 
However, our analysis \cite{Bednyakov:2015sca} demonstrated that in \cite{degrassi} 
theoretical uncertainty of the top-quark Yukawa coupling $y_t(M_t)$ 
seems to be underestimated by a factor of 2. 

\begin{wrapfigure}[19]{r}{0.5\textwidth}
	\centering
	\vspace{-10pt}
    \includegraphics[width=0.48\textwidth]{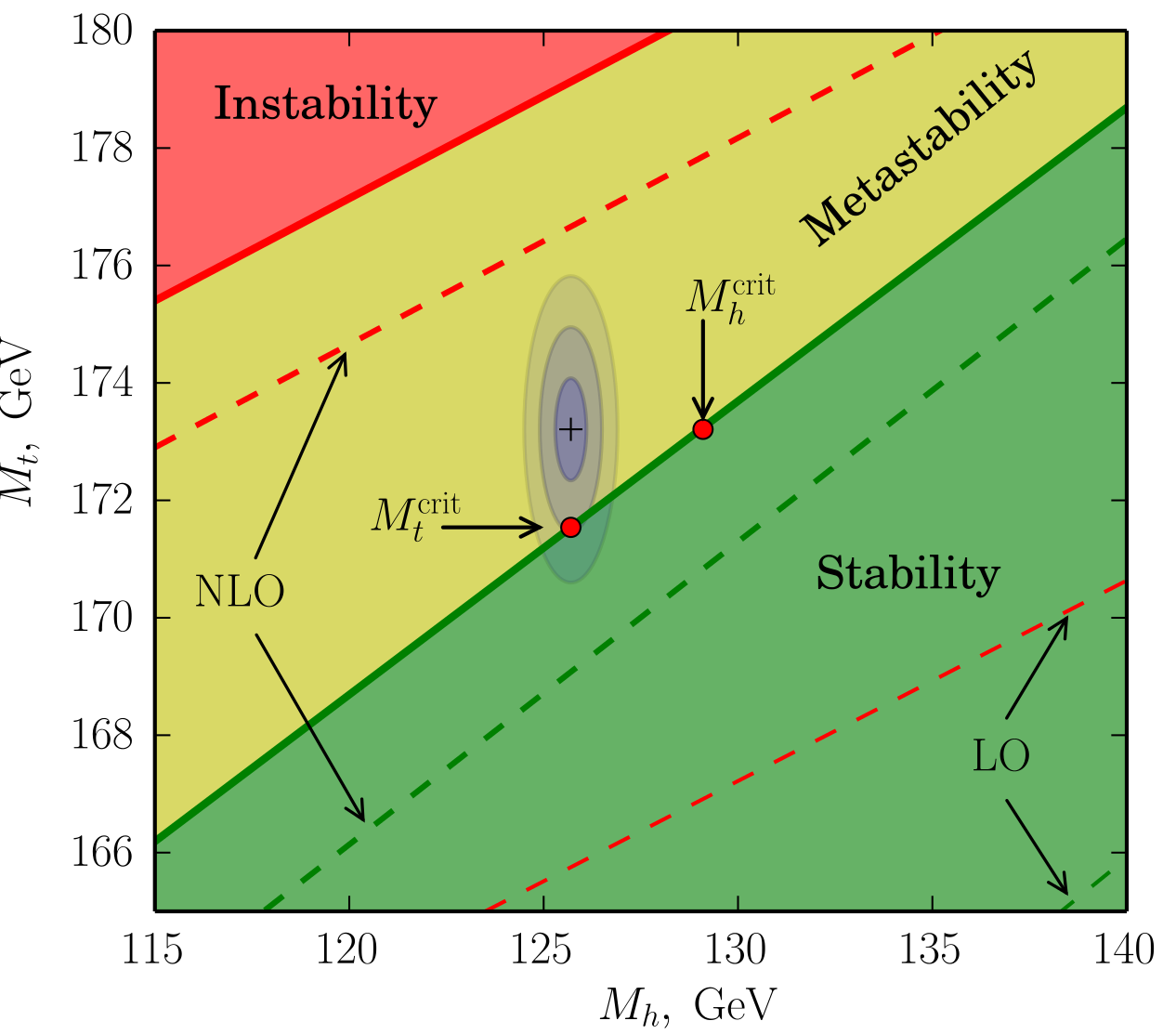}
	\vspace{-10pt}
  \captionof{figure}{Phase diagram of vacuum stability ontained at three loops (NNLO). 
	  One- (LO) and two-loop (NLO) results are also indicated.
	  The present world average of $(M_t, M_h)$ is shown with 1-3$\sigma$ countors.
  }
\label{fig:mhmt_phase}
\end{wrapfigure}

Let us now return to 
critical parameters and scales. 
Simple criterion of \eqref{eq:crit_cond_simple} serves as a good (and gauge-independent) starting point  
for our study. A more elaborated condition \eqref{eq:critical_conditions} requires evaluation of $\Veff$. 
In the latter case the prescription of Ref.~\cite{Andreassen:2014gha}
was utilized to compute $\Veff$ at the two-loop order.
According to this reference one not only improves the potential via RG, but also re-ogranize 
the expansion \eqref{eq:eff_pot_gen} by taking into account the scaling $\lambda \sim \hbar g_2^4$.
The detailed description of the procedure and the comparison of the simplified and full treatment of $\Veff$ can be found in Refs.~\cite{Andreassen:2014gha,Bednyakov:2015sca}.

A warning should be issued concerning physical meaning of ``critical'' (or ``instability'') scales emerging from the analysis. 
It is customary to think of these scales as scales, at which some New Physics might appear to cure potential instability.
However, one should be careful (see, e.g., discussions \cite{Espinosa:2015qea}) from associating the instability scale with Higgs field values, due to gauge-dependence of the latter.
We refrain from doing any conclusion on the New Physics scale and ignore this issue. 

The results of vacuum stability analysis are conventionally presented in a form of phase diagram (Fig.~\ref{fig:mhmt_phase}) in the plane $(M_h, M_t)$. 
Two critical lines correspond to the boundaries of the absolute stability region (green) and that of ``absolute'' instability (red) 
and are obtained in NNLO.
The importance of high-order corrections can be deduced from the comparison with the one-loop (LO) and two-loop (NLO) boundaries.
For physical masses lying in the red region (``Instability'') 
the EW vacuum decays \cite{Kobzarev:1974cp} 
during the age of the Universe $\tau_U$ with unit probability. The latter can be 
estimated from  
\begin{equation}
	\mathcal{P} \simeq \tau_U \left[ \frac{\tau_U^3}{R^4} \exp\left(-\frac{8\pi^2}{3|\lambda(1/R)|}\right) \right], 
	\quad
	\tau_U \simeq 13.8 \cdot 10^9~\mathrm{Years} 
	\simeq 6.6\cdot 10^{41}~\mathrm{GeV}^{-1}.
	\label{eq:decay_P}
\end{equation}
and is less than one in the yellow (``Metastability'') region.
In \eqref{eq:decay_P} $1/R$ corresponds 
the scale, at which $(-\lambda)$ is maximized, i.e., $\lambda(1/R)<0$ and $\beta_\lambda(1/R) = 0$.
The derivation of \eqref{eq:decay_P} is based on the simple approximation 
\eqref{eq:veff_approx} with negative $\lambda$ and, obviously, require more rigoruos, yet gauge-independent, 
treatment, e.g. along the lines of Ref.~\cite{Andreassen:2016cvx}. 
Nevetheless, it turns out that the measured values of $M_t$ and $M_h$ quoted in PDG2014~\cite{Agashe:2014kda} 
lie far away from the metastability boundary. 

In Fig.~\ref{fig:mhmt_phase} we also indicate our results for $\Mhcrit$ and $\Mtcrit$ 
\begin{equation}
	\Mtcrit = 171.54\pm0.30\pm 0.41~\GeV, \qquad \Mhcrit = 129.1\pm1.9\pm0.7~\GeV.
	\label{eq:crit_masses}
\end{equation}
The first error in \eqref{eq:crit_masses} corresponds to parameteric uncertainty due to 1$\sigma$ 
variation in the input parameters \cite{Agashe:2014kda}, while the second  --- to our conservative estimate of theoretical uncertainty coming from unknown high-order terms (see Ref.~\cite{Bednyakov:2015sca} for details).

	To summarize, the SM vacuum issue is studied at the three-loop order. The estimated theoretical uncertainties in critical parameters 
	are comparable with that due to the input, thus, requiring further improvement both from theory and experiment 
	(one anticipates the precision $\delta M_h = 0.04~\GeV$ and $\delta M_t = 0.1~\GeV$ at future linear colliders  
	\cite{Moortgat-Picka:2015yla}).
	Due to the fact that the pole mass is ill-defined for color particles, special attention should be paid to the top-quark mass parameter
	and the interpretation of the value quoted in PDG (see, e.g., \cite{Kieseler:2015jzh} and refs therein).
	\footnote{Alternative formulation when the stability constraint is emposed on $y_t$ can be found in Ref.~\cite{Bezrukov:2014ina}. }

	It is clear, however, that the whole analysis is carried out at zero temperature and 
	under assumption of the validity of the SM.
	Obviously, it can be modified by a bunch of factors, including gravity, finite-temperature effects 
	and, finally, New Physics. 
	It is very hard to review all the possibilitiese so we just refer here to some recent studies~\cite{
	NPstudies}. 
	In any case, the EW stability imposes imporant constraints on possible extensions of the SM. 
	Moreover, due to sucess of the SM, any New Physics should reproduce it as an effective theory at low energies.
	And last but not least, various cosmological implications 
	of the analysis can also be studied (see, e.g., \cite{Bezrukov:2014ipa,Espinosa:2015qea} and references therein).

	\section*{Acknowledgments}
	I am grateful to the organizers for the opportunity to give a talk on the conference
	dedicated to the 60th anniversary of my home institution - Joint Institute for Nuclear Research. 
	In addition, I would like to thank my colleagues, B.~Kniehl, A.~Pikelner, O.~Veretin and V.~Velizhanin, 
	for fruitful collaboration. Inspiring discussions with M.~Kalmykov and A.~Onischenko are kindly acknowledged.
	The work is supported in part by Heisenberg-Landau Programme and RFBR grant 14-02-00494-a.


\end{document}